\documentclass{jps-cp}
\usepackage{txfonts} 
\usepackage{wrapfig} 
\usepackage{graphicx} 
\usepackage{color}
\graphicspath{{figures/}}

\newcommand{\be}{\begin{equation}}
\newcommand{\ee}{\end{equation}} 
\newcommand{\beq}{\begin{eqnarray}}
\newcommand{\eeq}{\end{eqnarray}}

\def\D{\mathcal{D}}
\newcommand{\p}{\partial}

\newcommand{\bea}{\begin{eqnarray}}
\newcommand{\eea}{\end{eqnarray}}
\def\Tr{ \hbox{\rm Tr}}

\def\bra{\langle}
\def\ket{\rangle}
\def\half{\frac{1}{2}}

\def\e{\epsilon}

\def\half{\frac{1}{2}}
\def\dcfl{\Delta_{\textsc{cfl}}}

\def\SU{\rm{SU}}
\def\U{\rm{U}}
\def\bra{\langle}
\def\ket{\rangle}
\def \L{\textsc{l}}
\def \R{\textsc{r}}
\def \B{\textsc{b}}
\def \c{\textsc{c}}
\def \F{\textsc{f}}
\def \M{\textsc{m}}
\title{Stable non-Abelian semi-superfluid vortices in dense QCD}

\author{{Chandrasekhar Chatterjee}$^{a}$ and Muneto Nitta$^{ b}$}

\inst{Department of Physics, and Research and Education Center for Natural Sciences,\\ Keio University,Hiyoshi 4-1-1, Yokohama, Kanagawa 223-8521, Japan}

\email{chandra.chttrj@gmail.com$^{a}$, chandra@phys-h.keio.ac.jp$^{a}$, nitta(at)phys-h.keio.ac.jp$^{b}$}
 
\recdate{October 5, 2016}

\abst{Color superconductivity is expected to be formed in high density quark matter where color symmetry is spontaneously broken in the presence of  di-quark condensate. Stable non-Abelian vortices or color magnetic flux tubes  exist in the color-flavor locked phase at asymptotically high density.  $\mathbb{C}P^2$ Nambu-Goldstone(NG) bosons and Majorana fermions belonging to the triplet representation are  localized around a non-Abelian vortex.  We  discuss  the zero mode analysis and the low-energy effective  world sheet theory of a non-Abelian vortex. We determine the interactions of these bosonic and fermionic modes by using the nonlinear realization method. We also discuss the Aharanov-Bohm (AB) phases of charged particles, such as, electrons, muons, and color-flavor locked mesons made of tetra-quarks encircling around a non-Abelian vortex in the presence of electro-magnetic fields.
This is a review
based on our recent works
\cite{Chatterjee:2016ykq,Chatterjee:2015lbf,Chatterjee:2016tml}.
}

\kword{Color superconductor, CFL Phase, Non-Abelian vortices, Zero mode effective action, Majorana Fermion}

\begin{document}
\maketitle

\section{Introduction}
As it is well known  that quantum chromodynamics (QCD) describes dynamics of interacting quarks and gluons. However free quarks and gluons are not detected  in nature: they are confined to hadrons. 
The asymptotic freedom of $\SU(3)$ gauge theory describes QCD correctly at  high energies by perturbative quantum field theory, where the coupling constant is small.  At low energies the coupling constant becomes very strong and  the mechanism of confinement remains unsolved.  However if we increase the density sufficiently the system becomes asymptotically free and quarks condensate to color-superconductor due to existing attractive force among them.  At asymptotically high densities the mass of the strange quark can be neglected and the system reaches to the most symmetric color-flavor locked (CFL) phase \cite{Alford:1998mk,Alford:1997zt} 
(see Ref.~ \cite{ Alford:2007xm} as a review). 
 In this case,
the baryon number symmetry along with the $\SU(3)_{\c}$ color $\SU(3)_{\F}$ flavor symmetries are spontaneously broken by forming di-quark condensates. 
This creates a color superconductor, 
and by an analogy with ordinary metallic superconductor, 
one would expect the formation of vortices here. 
Because the $\U(1)_\B$ baryon number symmetry is a global symmetry, 
stable vortices which can be created in this medium is 
superfluid vortices.  
These vortices carry color magnetic fluxes, so we may call them as chromo-magnetic flux tubes. The situation is dual to confining flux tubes in hadronic phase where quarks are confined by chromo-electric flux tubes.  
Beside these  theoretical analogies, 
it is expected that a color superconductor can be found at the core of compact stars. The vortices in color superconductors then could effect the rotation dynamics of compact stars. 

In this talk,  based on 
our recent works
\cite{Chatterjee:2016ykq,Chatterjee:2015lbf,Chatterjee:2016tml},
we review some  developments of non-Abelian vortices 
in the CFL phase
after the comprehensive review paper 
\cite{Eto:2013hoa}. 
We discuss construction of vortices and their orientational zero modes using the Ginzburg-Landau (GL) formalism 
and Bogoliubov-de-Gennes (BdG) theory for fermions. 
We discuss interaction of fermion modes with bosonic orientational modes and write down effective interacting action. 
By introducing the electromagnetic interactions, 
we study AB phases of electrons, muons and CFL mesons 
around a vortex. 

 \section{Ginzburg-Landau free energy and non-Abelian vortices in the CFL phase}\label{sec:NAvor}
 \subsection{Ginzburg-Landau free energy}
Let us  first introduce  the GL description of color superconductors in 
the CFL phase.
 The GL order parameters are defined close to the critical temperature $T_{\rm c}$ by the di-quark condensates as
${\Phi_{\L}}_a^{\it A}  \sim  \e_{abc}\e^{\it ABC} {q_\textsc{l}}_b^{\it B} \mathcal{C}{q_\L}_c^{\it C}, 
\quad {\Phi_\R}_a^{\it A}  \sim  \e_{ abc}\e^{\it ABC} {q_\R}_b^{\it B} \mathcal{C}{q_\R}_c^{\it C},$
 where $q_{\L / \R}$ are left/right handed quarks carrying 
 fundamental color indices ${a, b, c}$ ($\SU(3)_{\c}$) and  fundamental flavour ($\SU(3)_{\L/\R}$) indices ${\it A, B, C}$. 
  At ground state, the chiral symmetry is spontaneously broken due to 
  $\Phi_\L = - \Phi_\R \equiv \Phi$. 
  The order parameter 
  $\Phi$ transforms as
$ \Phi' = e^{i\theta_\B}\U_\c \Phi \U_\F^{-1}, 
   \quad
 e^{i\theta_\B} \in \U(1)_\B, 
   \quad 
 \U_\c \in \SU(3)_\c,
   \quad 
 \U_\F \in \SU(3)_\F .$
After subtraction  of the redundant   discrete symmetries  the actual symmetry group is given by
${\rm G}  =  \scriptstyle
    \dfrac{\scriptstyle\SU(3)_{\c} \times \SU(3)_{\F} \times \U(1)_{\B}}
   {\scriptstyle\mathbb{Z}_3 \times \mathbb{Z}_3}.
\label{eq:sym_G}$

  The GL free energy  can be expressed as 
  \cite{Giannakis:2001wz,Iida:2000ha,Iida:2001pg}:
 \begin{align}
&&\textstyle \Omega =
\Tr\left[ {1\over 4 \lambda_3} F_{ij}^2 + {\varepsilon_3 \over 2} F_{0i}^2 
+ K_3\D_i \Phi^\dagger \D_i \Phi \right] 
+  \alpha \Tr\left(\Phi^\dagger \Phi \right)
+ \beta_1 \left[\Tr(\Phi^\dagger\Phi)\right]^2 
+\beta_2 \Tr \left[(\Phi^\dagger\Phi)^2\right]
\label{eq:gl}
\end{align}
where $i,j=1,2,3$ are space coordinates indices, $\lambda_{3}$ and $\varepsilon_{3}$  are 
the magnetic permeability and 
the dielectric constant for gluons, respectively.  
Here, $
\D_\mu \Phi = \p_\mu \Phi - i g_{\rm s} A^a_\mu T^a \Phi$ 
and 
$F_{\mu \nu} = \partial_{\mu}A_{\nu}
-\partial_{\nu}A_{\mu}-ig_{\rm s}[A_{\mu},A_{\nu}]$ are the covariant derivative and the field strength of gluons, respectively,
where $\mu,\nu$  indices are the  spacetime coordinates and
$g_{\rm s}$ is defined as the $\SU(3)_{\c}$ coupling constant. 
The microscopic calculations of the GL parameters $\alpha, \beta_{1,2}, K_3, \mu$ can be found in  Refs.~\cite{Giannakis:2001wz,Iida:2000ha}.

 The ground state is found to be 
$\bra \Phi \ket = \dcfl {\bf 1_3}$ where  
$\scriptstyle\dcfl \equiv \sqrt{-\frac{\alpha}{8\beta}}$,   
by which the full symmetry group $\rm G$ is spontaneously broken down 
to 
$\displaystyle{\rm{H} 
\simeq \dfrac{\scriptstyle\SU(3)_{\c+\F}}{\scriptstyle\mathbb{Z}_3}.}$
The order parameter space is found to be
$\displaystyle{{\rm G/H} \simeq 
{\SU(3) \times \U(1) \over {\mathbb Z}_3}
=\U(3)}$.

\subsection{Non-Abelian vortices, Nambu-Goldstone modes and Effective action}
We now review non-Abelian vortices in the CFL phase.   
The existence of vortices   are supported by  non-zero fundamental group $\pi_1 ({\rm G/H}) = \mathbb Z$.
 Here the vortices are global
(superfluid) vortices since the broken $\U(1)_{\B}$ is a global symmetry and  broken color symmetry generates chromo-magnetic fields inside the vortex
\cite{Balachandran:2005ev}. 
We can write down a particular vortex ansatz as 
 \begin{eqnarray}
 \label{colorvortexconfig1}
\Phi^0(r, \theta)  = 
\dcfl\left(
\begin{array}{ccc}
  e^{i\theta}f(r)& 0 & 0 \\
  0 &  g(r) & 0\\
  0 & 0 & g(r)
\end{array}
\right), \, \quad
A^0_i(r) = - \frac{1}{3g_s} \frac{\epsilon_{ij} x_j}{r^2} [1 - h(r)] \left(
\begin{array}{ccc}
 2 & 0 & 0 \\
 0 & - 1 & 0\\
 0 & 0& -1
\end{array}
\right). 
\end{eqnarray} 
The form of the profile functions $f(r)$ and $h(r)$ can be computed numerically with boundary condition,
$ f(0) = 0, \quad 
\p_r g(r)|_0 = 0, \quad 
h(0) = 1,   \quad f(\infty) = g(\infty) = \dcfl,  \quad 
h(\infty) = 0 .$

As it can be shown from the boundary condition of the profile functions that the vortex configuration in 
Eq.~(\ref{colorvortexconfig1})  breaks spontaneously the unbroken ground state
symmetry group $\SU(3)_{\rm{\c+\F}}$ 
into a subgroup $\SU(2)\times \U(1)$  
inside the vortex core. 
This symmetry breaking generates continuous degeneracies in vortex solutions and  the corresponding NG modes are parametrized by 
the coset space 
$\scriptstyle\frac{\SU(3)}{\SU(2)\times \U(1)} \simeq {\mathbb C}P^{2}$ 
\cite{Nakano:2007dr}.
   Applying a global transformation by 
a reducing matrix 
$\scriptstyle {\U} = 
\frac{1}{\sqrt{X}}\left(
\begin{array}{cc}
 \scriptstyle 1 & \scriptstyle -B^\dagger \\
\scriptstyle B & \scriptstyle X^\half Y^{-\half} 
\end{array}
\right), \quad 
X = 1 + B^\dagger B, \quad Y = {\bf 1}_3 + BB^\dagger, $
we can generate the generic solutions on the ${\mathbb C}P^{2}$ space, where $B = \{B_1, B_2\}$ are inhomogeneous 
coordinates of  ${\mathbb C}P^{2}$:
\begin{equation}
 \label{colorvortexconfigcp2}
\Phi(r, \theta)  = U \Phi^0(r, \theta) U^\dagger
,\quad
A_i(r) = U A^0_i(r) U^\dagger .
\end{equation}
The low-energy excitation and interaction of these NG zero modes are described by
 the effective  action  as
the $\mathbb{C}P^2$ nonlinear sigma model on the world-sheet 
coordinate $(t, z)$ as
\cite{Eto:2009bh,Eto:2009tr,Chatterjee:2016tml}:
\begin{eqnarray}
\mathcal{L}_{\mathbb{C}P^2} =  \mathcal{C}_\alpha \left[ |\p_\alpha \hat n|^2 +  (\hat n ^\dagger \p_\alpha \hat n)^2\right], \qquad
\hat n^T = \frac{1}{\sqrt{1 + |B_i|^2}}
\left(1,  B_1,  B_2\right)
\end{eqnarray}
where $\mathcal{C}_\alpha$ are constants 
written as the integrations of profile functions.
This result was obtained first by taking 
a singular gauge
\cite{Eto:2009bh,Eto:2009tr} 
and has been recently confirmed also in a regular gauge
\cite{Chatterjee:2016tml}.

\section{Fermion zero modes and effective action}
\subsection{Fermion zero modes in a particular vortex}
To compute fermion zero modes in a particular vortex background we start with the BdG Hamiltonian
\begin{eqnarray}
\label{H}
{\cal H} \, = \, \bar\Psi_a^A\, \left(\hat{\cal H}_0\,\delta_{ab} \, \delta^{AB}  \, + \, \tilde \Phi^{AB}_{ab} \, \right) \, \Psi_b^B,\quad \Psi = 
\left(
\begin{array}{c}
 \psi  \\
  \psi_c   
\end{array}
\right),\quad \psi_a^A=(\psi)_{A a}= 
\left(
\begin{array}{ccc}
s_b  \,& s_g\, & s_r\\
d_b \,& d_g \,& d_r\\
u_b \,& u_g \,& u_r
\end{array}
\right)
\end{eqnarray}
 where $ a, b = \{b, g, r\}, \{A, B\} =  \{s, d, u\}$.   $\Psi$ is the quark quasi-particle  field written in the Nambu-Gor'kov basis and $\psi_c \, = \,  e^{i\eta_c} i \gamma^2 \,  \psi^*$ where $\eta_c$ is an arbitrary phase.  Here $\psi$ is written as a $3\times 3$ matrix whose entries are  quarks with color and flavor indices.
  $ \hat{\cal H}_0$ and $\tilde \Phi^{AB}_{ab} $ can be expressed as
 \begin{eqnarray}
 \hat{\cal H}_0 = 
\left(
\begin{array}{cc}
 -i \vec\gamma \cdot \vec\nabla - \gamma^0 \mu & 0  \\
 0  &    -i \vec\gamma \cdot \vec\nabla  +  \gamma^0 \mu\\
\end{array}
\right), \qquad
\tilde \Phi^{AB}_{ab} = 
\left(
\begin{array}{cc}
0  &  \Phi^{AB}_{ab} \, \gamma^5  \\
 - {\Phi^*}^{AB}_{ab} \,  \gamma^5 &   0  \end{array}
\right),
\end{eqnarray}
where  $\Phi^{AB}_{ab}$ is the generic gap function of a particular vortex defined as $\Phi^{AB}_{ab}\sim \epsilon_{abc}\epsilon^{ABC}\Phi^C_c$ with $\Phi$ defined in Eq.~(\ref{colorvortexconfig1}).
  The BdG eigenvalue equation is found to be, 
${\cal H}\Psi = {\cal E} \Psi .$
A  normalizable triplet zero mode can be shown to exist by solving BdG equation and the equations for the triplet states are found to be
\begin{align}
\left(\hat{\cal H}_0 -  \tilde\Delta_1\right)\Psi^1 = 0,\quad \left(\hat{\cal H}_0 +  \tilde\Delta_1\right)\Psi^2 = 0,\quad  \left(\hat{\cal H}_0 -  \tilde\Delta_1\right)\Psi^3 = 0, 
\end{align}
where $\scriptstyle\psi^1 = \frac{d_r + u_g}{\sqrt 2}, \quad \psi^2 =  \frac{d_r - u_g}{i\sqrt 2}, \quad  \psi^3 = \frac{u_r - d_g}{\sqrt 2}$ and $\scriptscriptstyle\tilde\Delta_1 = \scriptscriptstyle
\dcfl\left(
\begin{array}{cc}
\scriptscriptstyle 0  & \scriptscriptstyle  f(r) e^{i\theta} \, \gamma^5  \\
 \scriptscriptstyle - f(r) e^{-i\theta} \,  \gamma^5 &  \scriptscriptstyle 0  \end{array}
\right) $. The  triplet zero modes were found in Refs.~\cite{Yasui:2010yw,Fujiwara:2011za}.
To write down the effective action we introduce the $t, z$  dependence  in a factorized way as:
$\Psi_L^\rho(t, z, x, y) = \chi_L^\rho(t, z) {\Psi_0}_L^\rho(x, y),$ $ \Psi_R^\rho(t, z, x, y) = \chi_R^\rho(t, z) {\Psi_0}_R^\rho(x, y)$
where $\rho = \{ 1, 2,3\}$(no summation over $\rho$). Inserting this into the original action  we may write down the effective action as
\begin{equation}
{\cal L}_{\rm eff} 
= -i \Tr \,\chi^\dagger\left( \dot\chi - v_{\rm fermi}\sigma^3 \p_z \chi\right), \quad \chi = \chi^\rho \tau^\rho,\quad \tau^\rho = \half \sigma^\rho
\end{equation}
where $v_{\rm fermi}$ is the velocity and the two-dimensional spinors ${\chi}^{\rho}(t, z)$ are defined by
$\scriptstyle{\chi}^\rho(t, z) = \scriptstyle
\left(
\begin{array}{ccc}
\scriptstyle {\chi}_L^\rho(t, z)  \\
\scriptstyle {\chi}_R^\rho(t, z) \end{array}
\right).$

\subsection{The interaction between NG mode and fermion zero mode: nonlinear realization}
So far we discussed NG modes and fermion zero modes in the background of a particular vortex solution along the vortex configuration at $B_i =0$ separately. However they interact once we excite the NG modes ($B_i$) in the presence of 
fermion zero modes. This can be understood as follows. The fermion modes are computed in the particular vortex background along the vortex configuration at $B_i =0$.
At that point on  $\mathbb{C}P^2$ moduli space the fermion zero modes transform under unbroken SU$(2)$ as triplet linearly. However when we excite the NG modes
along a path on the $\mathbb{C}P^2$ moduli space then the transformation matrices of the fermion  zero mode changes since the background vortex configuration changes along the path on moduli space.  So the transformation becomes a nonlinear action of the full group SU$(3)$ on the fermion
modes. Since NG modes($B_i(t, z)$) are functions of the $t,z$ coordinates, the transformation becomes local on the $t-z$  plane and it generates gauge fields which is basically 
projection of pure gauge along the SU$(2)$ subspace.
\begin{eqnarray}
A_\alpha = -i\left[U^\dagger(\vec B(t, z))\partial_\alpha U(\vec B(t, z))\right]_{\perp}, \quad\alpha = \{0, 3\}
\end{eqnarray}
here $\perp$ shows the projection along  unbroken $\SU(2)$. The effective action can be written as,
\begin{eqnarray}
\label{LEff}
{\mathcal L}_{\rm fermi} = -i {\rm Tr} \,\left[\chi^\dagger(t, z)\left\{ \D_0\chi(t, z) - v_{\rm fermi}\sigma^3\D_z \chi(t, z)\right\}\right] ,
\end{eqnarray}
where $v$ is 
the same for all three triplet zero modes \cite{Chatterjee:2016ykq}.

\section{Electromagnetic Interaction and Aharonov-Bohm(AB) Effect}
So far we have neglected 
the electromagnetic (EM) interaction. 
Here we introduce the EM interaction ($\U(1)_{\rm{em}}$)  as a part of 
the flavour symmetry group $\SU(3)_{\F}$ and the generator of $\U(1)_{\rm{em}}$ is defined as: 
 \begin{eqnarray}
 \label{emcharge}
Q  = \frac{1}{3}
\left(
\begin{array}{ccc}
 2 &  0  & 0  \\
 0 & -1  & 0  \\
 0  &  0 &   -1
\end{array}
\right).
\end{eqnarray}  
The introduction of the 
 EM interaction brings two effects. 
 One is that it makes some of ${\mathbb C}P^2$ modes massive \cite{Vinci:2012mc} and the other is the effective ${\mathbb C}P^2$ model 
 is gauged by which the localized ${\mathbb C}P^2$ modes interact with the bulk EM fields \cite{Hirono:2012ki}.
In this case the massive and massless diagonal gauge fields in the bulk 
can be expressed  as 
\begin{eqnarray}
&& A^{\M}_\mu = \frac{g_s}{g_\M} A^8_\mu - \frac{\eta e}{g_{\M}} A^{\rm{em}}_\mu, 
\quad
A^q_\mu = \frac{\eta e}{g_\M}A^8_\mu + \frac{g_s}{g_\M}A^{\rm{em}}_\mu ,
\end{eqnarray}
respectively, 
where $\scriptstyle\eta = \frac{2}{\sqrt 3}$ and $ g_\M^2 = g_s^2 + \eta^2 e^2 $.  
So the EM interaction is effectively generated by $A^q$ gauge field since all fields living in the bulk interact only with $A^q$ effectively. Here we may define the effective EM group 
as $\tilde\U(1)^{\rm{em}}$. The original EM gauge potential can be written as 
$
A^{\rm{em}}_\mu = \frac{g_s}{g_\M} A^q_\mu - \frac{\eta e}{g_\M} A_\mu^{\M}
$, and when in the bulk the massive part vanishes, the effective EM coupling of a particle with charge $q$ becomes 
$\textstyle\frac{q g_s}{\sqrt{g_s^2 + \eta^2 e^2}}.
$ 
In the presence of the EM interaction, the $\mathbb{C}P^2$ moduli space is reduced to $\mathbb{C}P^1$ with one extra point on moduli space. The point is labeled  as a Balachandran-Digal-Matsuura(BDM)  vortex \cite{Balachandran:2005ev}, which is given by Eq.~(\ref{colorvortexconfig1}). The ansatz for one of the $\mathbb{C}P^1$ vortex can be expressed as \cite{Eto:2009tr}
\begin{align}
\Phi(r, \theta)  =  
    \dcfl\left(
\begin{array}{ccc}
 g(r) &  0  &  0  \\
  0&  e^{i\theta}f(r)& 0\\
  0 &  0 &  g(r) 
\end{array}
\right) , &\quad
 A^{\M}_i(r){ \rm T}^8 =  \frac{1}{6 g_{\M}} \frac{\epsilon_{ij }x_j}{r^2} [1 - h(r)] \left(
\begin{array}{ccc}
 2 &  0 & 0 \\
 0 & - 1 &  0\\
 0 & 0&  -1
\end{array}
\right) , \nonumber\\ 
A^3_i(r) {\rm T}^3 = - \frac{1}{2g_s} \frac{\epsilon_{ij }x_j}{r^2} [1 - & h(r)]\left(
\begin{array}{ccc}
 0\, &\, 0 &  0 \\
 0\, & \,  1 &  0\\
 0\, &\,  0&  -1
\end{array}
\right). 
\end{align}
As it is discussed above that the effective charge of a charged particle becomes fractional in the bulk. So when a charged particle encircles any vortex in bulk, it will pick up a non-trivial AB phase. AB phase of any particle with charge $e$ can be written in this case as
$ \varphi_{\rm AB} =  \frac{\eta e^2}{2 \pi g_\M}\left| \oint A^{\M}\cdot dl\right|.$
The AB phases  of electrons, muons and charged CFL mesons due to BDM and $\mathbb{C}P^1$ vortices are computed as\cite{Chatterjee:2015lbf}
\begin{eqnarray}
\label{ABbdm}
\varphi^{\textsc{bdm}}_{\rm AB} =  \frac{\eta e^2}{2 \pi g_\M} \left|\oint A^{\M}\cdot dl\right| 
=  \frac{2 e^2}{3 g_s^2 + 2 e^2},\quad
\varphi^{\mathbb{C}P^1}_{\rm AB} =  \frac{\eta e^2}{2 \pi g_\M}\left| \oint A^{\M}\cdot dl \right| =  \frac{e^2}{3 g_s^2 + 2 e^2}.
\end{eqnarray}

\section{Summary and Discussion}
In this talk, we have discussed non-Abelian semi-superfluid vortices in 
the CFL phase of dense quark matter.  
After introducing bosonic and fermionic zero modes around a single vortex,
we have written down the interacting effective action of these modes around the vortex. 
After introducing the EM interaction we have written down AB phases of bulk light particles such as electrons, muons, CFL mesons. 

Recent topics 
after the review
\cite{Eto:2013hoa}
which this review cannot cover 
includes 
decay of an Abelian vortex into three non-Abelian vortices
\cite{Alford:2016dco},
a non-Abelian vortex lattice and a color magnetism 
\cite{Kobayashi:2013axa},  
topological superconductivity \cite{Zubkov:2016llc}. 

The hadron-quark duality between the confinement and CFL phases
was studied in the framework of the GL theory \cite{Eto:2011mk}.
Introducing fermion zero modes as discussed may change the situation.
The effect of AB phases 
in the transport of bulk particles in the CFL phase
can be computed as was done in the 2SC phase 
\cite{Alford:2010qf}.
\section*{Acknowledgement}
This work is supported by the Ministry of Education, Culture, Sports, Science (MEXT)-Supported Program for the Strategic Research Foundation at Private Universities ``Topological Science'' (Grant No. S1511006). 
C.~C. acknowledges support as an International Research Fellow of the Japan Society for the Promotion of Science (JSPS). 
The work of M.~N. is supported in part by a
JSPS Grant-in-Aid for Scientific Research (KAKENHI Grant No. 16H03984)
and by a Grant-in-Aid for
Scientific Research on Innovative Areas ``Topological Materials
Science'' (KAKENHI Grant No.~15H05855) and ``Nuclear Matter in Neutron
Stars Investigated by Experiments and Astronomical Observations''
(KAKENHI Grant No.~15H00841) from the the Ministry of Education,
Culture, Sports, Science (MEXT) of Japan.

\end{document}